# Observation of hedgehog skyrmions in sub-100 nm soft magnetic nanodots

**Eider Berganza[1*], Miriam Jaafar[1*], Maite Goiriena-Goikoetxea[2,3], Javier Pablo-Navarro[4], Alfredo García-Arribas[3,5], Konstantin Gusliyenko[6,7], Cesar Magén[4,8,9], José María de Teresa[4,8,9], Oksana Chubykalo-Fesenko[1] and Agustina Asenjo[1]**

[1] Instituto de Ciencia de Materiales de Madrid, CSIC, 28049 Madrid, Spain
[2] Department of Electrical Engineering and Computer Science, University of California, Berkeley, CA 94720, USA.
[3] Department of Electricity and Electronics, University of the Basque Country (UPV/EHU), 48940 Leioa, Spain.
[4] Laboratorio de Microscopías Avanzadas (LMA) - Instituto de Nanociencia de Aragón (INA), Universidad de Zaragoza, 50018 Zaragoza, Spain.
[5] Basque Center for Materials, Applications and Nanostructures (BCMaterials), Parque Tecnológico de Bizkaia, Building 500, 48160 Derio, Spain.
[6] Department of Materials Physics, University of the Basque Country (UPV/EHU), 20018 Donostia, Spain
[7] IKERBASQUE, the Basque Foundation for Science, 48013 Bilbao, Spain
[8] Instituto de Ciencia de Materiales de Aragón (ICMA), Universidad de Zaragoza-CSIC, 50009 Zaragoza, Spain.
[9] Departamento de Física de la Materia Condensada, Universidad de Zaragoza, 50009 Zaragoza, Spain.

**Magnetic skyrmions are nanometric spin textures of outstanding potential for spintronic applications due to unique features governed by their non-trivial topology. It is well known that skyrmions of definite chirality are stabilized by the Dzyaloshinskii-Moriya exchange interaction (DMI) in bulk non-centrosimmetric materials or ultrathin films with strong spin-orbit coupling in the interface. In this work, we report on the detection of magnetic hedgehog-skyrmions at room temperature in confined systems with neither DMI nor perpendicular magnetic anisotropy. We show that soft magnetic (permalloy) nanodots are able to host non- chiral hedgehog skyrmions that can be further stabilized by the magnetic field arising from the Magnetic Force Microscopy probe. Analytical calculations and micromagnetic simulations confirmed the existence of metastable Néel skyrmions in permalloy nanodots even without external stimuli in a certain size range. Our work implies the existence of a new degree of freedom to create and manipulate skyrmions in soft nanodots. The stabilization of skyrmions in soft magnetic materials opens a possibility to study the skymion magnetization dynamics otherwise limited due to the large damping constant coming from the high spin-orbit coupling in materials with high magnetic anisotropy.**

Magnetic skyrmions, topologically protected inhomogeneous spin textures on the nanoscale, named after T. Skyrme who predicted similar non-linear configurations in quantum field theory,[1] are widely investigated due to their potential applications in memory storage devices, spintronics as well as due to their unusual fundamental properties.

With their small sizes and high mobility under low current densities, skyrmions are envisaged as information carriers in the racetrack memory type devices. In this sense, control over the formation and manipulation[2,3] of skyrmions, particularly in confined geometries,[4,5] becomes of uttermost importance.

Most of the works in the literature report on skyrmions in systems with broken inversion symmetry allowing the Dzyaloshinskii–Moriya exchange interaction[6] (DMI) either in ultra-thin multilayers of transition metals and materials with strong spin-orbit coupling[7,8,9] or in non-centrosymmetric compounds.[10] Both Néel (hedgehog) and Bloch skyrmions can be stabilized in the above mentioned cases due to the interplay of DMI and uniaxial magnetic anisotropy.

Non-chiral Bloch skyrmions (or bubble domains) can be stabilized in infinite films with out-of-plane magnetic anisotropy by the dipolar interactions without the need of DMI. A separate question is the stabilization of Bloch skyrmions in nanostructured materials such as nanodots where the dipolar interactions play a very important role and change the skyrmion stability conditions.[11] Up to now, in all cases the perpendicular magnetic anisotropy was considered a necessary ingredient. Particularly, soft magnetic permalloy (Py, NiFe alloy) dots are believed to host magnetic vortices only[12,13] (half Bloch skyrmions). Nobody so far assumed that stabilization of the Néel skyrmions in confined systems without magnetic anisotropy was possible.

On the other hand, essential progress of the experimental techniques during the last decade enabled the detection the magnetic vortices,[14] bubbles[15] or skyrmions[16] by using various imaging techniques on the nanoscale. The importance of imaging individual nano-objects lies in its capability to directly visualize these spin textures or even to control them. Several techniques such as Photo Emission Electron Microscopy (PEEM)[17] or Electron Holography (EH)[5] are commonly used to study magnetic configurations of individual nano-objects. Nevertheless, the family of Scanning Probe Microscopy (Spin Polarized Scanning Tunneling Microscopy, SP-STM,[3] Magnetic Exchange Force Microscopy, MExFM[18] and Magnetic Force Microscopy,[19,20] MFM) provides higher resolution images as well as remarkable sensitivity.

In this work, we studied permalloy (Py) sub-100 nm diameter particles (nanodots) with no perpendicular uniaxial magnetic anisotropy or DMI. The expected magnetization configuration for Py nanodots in this size range is either a vortex state or a single domain state (with in-plane magnetization) due to the low magnetocrystalline anisotropy.[13,21] However, Néel (hedgehog) skyrmion spin textures were unexpectedly detected in the present study through MFM. The magnetization configuration of the Néel skyrmions[22,23] is characterized by spins rotating in radial planes from their cores to the boundaries, in contrast to Bloch skyrmions, where the spins are essentially perpendicular to the radial planes, forming closed-flux circles. The evolution of the magnetic configuration under external applied in-plane (IP) fields leaves no doubt of the existence of radial hedgehog skyrmions in such Py dots. Analytical calculations showed that these magnetization textures are metastable high-energetic states and micromagnetic simulations confirmed this fact.

## Detection of skyrmions

For the present experiments, Py nanodots with a diameter of 70 nm and height of 30 nm were grown by hole mask colloidal lithography (HCL) as explained elsewhere.[24] The nanodots studied along this investigation are distributed onto a silicon substrate far enough from neighboring nanodots in such a way that they can be considered non-interacting with each other. Their shape and size were assessed by Scanning Electron Microscopy (SEM) and High Resolution Transmission Electron Microscopy (HR-TEM) which confirm the polycrystalline nature and therefore, the absence of magnetocrystalline anisotropy. The shape of the nanodots is close to a hemisphere

MFM imaging was performed on the sample (hereinafter referred to as A) using a commercial CoCr coated tip from Nanosensors[TM]. The experiments were performed in remnant magnetization state, after applying a saturating magnetic field to the sample in the out-of-plane (OOP) direction to promote the vortex formation.

The MFM contrast observed in these experiments (see Figure 1a and 1b) resembles a vortex magnetization configuration with a core distinguished from the surrounding area. Vortices present IP closed-flux magnetization with a core at the centre, where the OOP magnetization component can be either positive or negative and their chirality can be clock or anticlock-wise[25,26,27]. However, the strong dark contrast around the core and the fact that all the cores present bright contrast, regardless of the tip polarization (see Figure 1e), lead us to explore alternative possible magnetization configurations.

In the MFM image of a vortex (Figure 1d, Py nanodiscs of 140 nm in diameter), the core is distinguished from the surrounding area due to the high magnetostatic interaction between the tip stray field and the magnetic moment of the core. Dark (or bright) core contrast corresponds to attractive (or repulsive) interaction, *i.e.*, to the core magnetization parallel (or antiparallel) to the tip polarization (Figure 1c). Thus, the core MFM contrast can be reversed by switching either the magnetization of the tip (Figure 1d) or the polarity of the core. However, in sub-100 nm nanodots the contrast at the center of the nanoparticles is always positive (corresponding to an antiparallel tip-core configuration, as in Figure 1e), regardless of the MFM probe polarization.

The fingerprint of the vortex spin texture is the perpendicular displacement of its core with respect to an IP applied magnetic field, until the critical field is reached and the magnetic moments completely align with the in-plane field. It is well known that the core displacement direction depends entirely on the chirality and it is therefore independent of its magnetization direction.[28,29] Nonetheless, Variable Field (VF)-MFM[30] results show that in the sub-100 nm nanodots the displacement of the core is either parallel or antiparallel to the in-plane magnetic field (see sequence I and II in Figure 2). This is indicative of the absence of IP closed-flux magnetization configuration. Notice that the magnetic state of the tip has been defined prior to the measurements (Figures 2a and 2e). As evidenced in Figure 2a, the white core is moving antiparallel to the applied field (sequence I). This confirms the hypothesis that the dot magnetization configuration is not a vortex, but rather a hedgehog skyrmionic configuration with an essential radial IP magnetization component. Néel (hedgehog) skyrmions[23] are characterized by a rotating OOP magnetization component from the core to the dot edges by means of radial curling of the spins. The cores of these configurations displace parallel or antiparallel to the in-plane magnetic field direction increasing size of the domain with magnetization parallel to the field.

The core displacement when IP magnetic field is applied depends solely on the skyrmion chirality *sign ($m_\phi$)* for Bloch skyrmions (vortices) and on the radial component direction *sign ($m_\rho$)* for Néel skyrmions, considering a unit cylindrical magnetization vector **m** *($m_\rho$, $m_\phi$, $m_z$)* with the z axis directed perpendicularly to the film surface. If an in-plane bias field is directed along *Oy* axis *($H_y$ >0)*, such as in Figure 2a (and Figure 2e), the skyrmion core moves along the Oy axis. Thus, the Néel skyrmion core displacement Y fulfils the rule: *sign (Y) = - sign ($m_\rho$)*.

In a hemispherical dot *(z > 0)* or a truncated hemisphere *(z > $z_c$ >0)*, the radial magnetization component and the skyrmion core polarity $p = sign (m_z (0))$ are not independent. The core polarity sign determines the radial magnetization direction, namely from inside-out or from outside-in, respectively. Thus, in sequence I ($p > 0$) the core displaces antiparallel to the field, owing to the radial magnetization configuration pointing outwards (Figure 2c). In sequence II ($p < 0$) the skyrmion core displaces parallel to the field as it is indeed observed in the experiment, since the radial magnetization points inwards.[31]

At this point, it is important to notice that the two configurations of skyrmions observed in sequence I and II have been imaged under presence of the local magnetic stray field emerging from the tip. The stray field of the tip plays a crucial role on the definition of the magnetic configuration of the nanodots, considering that a positive or negative core skyrmion can be induced by applying a perpendicular field in the appropriate sense. Therefore, the fact that an OOP local field determines the development of a positive or negative core skyrmion configuration explains why the images obtained in the MFM measurements (*i.e.*, white core and dark contour) remain the same regardless of the magnetization direction of the tip. Simulated MFM images shown in Figure 2d reproduce nicely the experimental data.

Permalloy hemispherical nanoparticles were modelled via micromagnetic simulations with initial conditions close to that of the Néel skyrmion. Simulations in Figure 2b and Figure 2f reproduce the skyrmion as a minimum energy state and prove that a hedgehog skyrmion configuration is, at least, metastable in hemispheres of 60 nm diameter. They demonstrate the stability of both positive (and negative) polarity hedgehog skyrmions, with the radial component of the magnetization pointing out (and in), in good agreement with the observations. Results unveil that the core diameter is wide at the base and it narrows approaching the nanodot surface. Additionally, when it is subjected to external field, the magnetic moments of the base respond later than the ones closer to the dot upper surface. Notice that the out-of-plane ($m_z$) magnetization component is represented in a color scale and that a transversal dot cross-section is shown to emphasize the movement of the core. Images in Figure 2c display the base plane of the nanodot with radial configuration ($m_y$ depicted in a blue colour scale), which is present when the skyrmion has not yet been annihilated.

## Skyrmion stability assessment

Analytical calculations, carried out to shed light on the experimental result, predict the existence of a hedgehog skyrmion in Py nanoparticles with no perpendicular anisotropy or DMI (see Figure 3a). Yet, only vortices or single domain states have been reported so far in these soft magnetic structures in the literature.[21] Analytical calculations were conducted, in parallel to micromagnetic simulations, with the aim to verify the stability of the Néel skyrmions in permalloy nanodots in absence of external fields. Our calculations show that Néel skyrmions indeed exist in permalloy dots as high-energy metastable states, *i.e.*, with higher energy than Bloch skyrmion (or vortex) states. In Figure 3b, the reduced skyrmion energy is plotted versus the dot aspect ratio, particularized for the case $R/l_{ex}=10$. $l_{ex}=\sqrt{2A/\mu_0 M_s^2}$ is the material exchange length, A the exchange parameter and $M_s$ accounts for the saturation magnetization.

These Néel skyrmion states exist in some regions in the phase diagram (regions delimited by the black dots and lines in Figure 3a) and become inaccessible for large dot thicknesses L, or small dot radius R, due to their very high energies. Owing to the increasing magnetostatic energy when the ratio L/R increases, the upper boundary in Figure 3a presents a downwards trend. On the other hand, the exchange energy increases with the dot radius *R* decreasing at fixed L/R, making the Néel skyrmion unstable at small *R* (left boundary in Figure 3a marked with black squares). Results obtained through micromagnetic simulations on semispherical dots have been added to the graph in Figure 3a. The regions where the IP single domain and the vortex states are ground states in the system are shown in light blue and grey regions, respectively. Moreover, the graph indicates the geometries where the metastable skyrmion states were found (see skyrmions represented as red-blue dots). Note that we did not find stable skyrmions for dots with radius below 30 nm in hemispheres.

According to micromagnetic simulations, the Néel skyrmion states are highly metastable and are destroyed by small perturbations. Experimentally, we have verified that they exist over larger interval of the applied field values which points towards an additional stabilization by the stray field coming from the MFM tip.[32] Further experiments were carried out to assess the influence of the localized magnetic field of the tip on the skyrmion stability. With this purpose, a low moment MFM tip was designed for comparison. The tip magnetic stray field can be tuned simply by controlling the amount of the magnetic coating (usually CoCr) deposited.[33] However, this method has some limitations; as the coating thickness decreases the MFM signal to noise ratio can be jeopardized. In this work, Fe nanorods[34] were grown by Focused Electron Beam Induced Deposition (FEBID) on standard AFM tips.[35] Nanorod diameters below 35 nm with very sharp endings (7 nm) can be obtained to improve the resolution. Moreover, the cylindrical geometry and the high aspect ratio of the Fe nanorods maximize the OOP vs. IP stray field ratio. To obtain tips with intermediate stray fields, Co was sputtered onto commercial AFM tips choosing the desired thickness (see Methods), *i.e.* adjusting the local stray field. Several tips of categories 1-3 were fabricated for this study (details are gathered in Table 1).

| Tip category | 1 | 2 | 3 |
|---|---|---|---|
| | 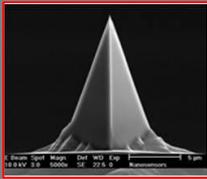 | 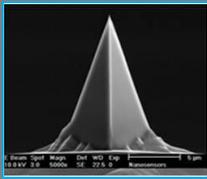 | 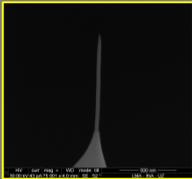 |
| Type | Nanosensors commercial | Sputtering cobalt coating | Fe nanorod |
| Properties | Coating: 50-25 nm | Coating: 35-20 nm | Diameter: 30-50 nm<br>Length: 1007 nm |
| OOP stray field | 29-24 Hz | 19-15Hz | 22-4Hz |

**Table 1. Description of the MFM probes.** Features of the tips utilized for the VF-MFM experiments summarized in Figure 4.

Making use of series of VF-MFM based methods,[36] the critical fields of the skyrmions have been determined. In Figure 4a, the skyrmion existence field ranges — field ranges between nucleation and complete annihilation- are depicted as a function of the stray field of the probes. In Supplementary Information IV, an example was introduced to explain how the critical fields were measured, together with the method used for the determination of the tip stray fields.

Figure 4b-d display representative standard MFM of skyrmions under in-situ applied fields, performed with probes of different stray field values. The the color code of frames of MFM images is in accordance with that shown in Table 1. Interestingly, the skyrmion annihilation field is almost 5 times larger when a commercial probe is used (tip 1, Figure 4b) in comparison with the values obtained for the nanorod probe (tip 3, Figure 4d). Thus, larger OOP stray fields lead to the stabilization of the skyrmion, turning it into a magnetically harder configuration. As in previous works where the local field of the tip is used to manipulate the magnetic charges,[37,38] in this work the tip stray field serves as a tool to control the stability of the skyrmion configuration.

The application of a local out-of-plane stray field, can act as a modulator to tune the stability of the skyrmion, confined in a soft nanodot. Generally speaking, the stabilization of hedgehog skyrmions in a material with neither DMI interaction nor magnetic anisotropy invites to lead efforts to span the range of magnetic systems where topologically non trivial spin textures can be found.

Additionally, Py nanodots hosting a Néel skyrmion, might be particularly interesting for the study of skyrmion-based spin-torque nano oscillators (STNO),[39,40] which according to predictions would have no threshold current for activation, in contrast to vortex-based STNO. The use of a suitable soft magnetic material, with high saturation magnetization, low anisotropy and its well-defined magnetic resonance spectra, opens a path to its application in microwave devices based on the fast magnetization dynamics.

## Conclusions:

In summary, by using magnetic force microscopy techniques we detected for the first time hedgehog (Néel) skyrmions in soft magnetic Py nanoparticles. This is a non-expected magnetization configuration in the nanoparticles without DMI and more remarkably, in absence of any magnetic anisotropy. We demonstrated theoretically and experimentally that the Néel skyrmions exist in permalloy circular nanodots as highly metastable states and that they can be stabilized by the local field coming from the MFM tip. Under in-plane applied magnetic field, the hedgehog skyrmions behave as it is theoretically predicted, *i.e.* their core moves parallel or antiparallel to the field in agreement with its core polarization direction. The skyrmion stability is studied as a function of an out-of-plane local field. We believe that the existence of the Néel skyrmions in nanodots without magnetic anisotropy and negligible DMI opens new perspectives for the exploration of topologically non-trivial spin textures in systems that have not been yet considered. Moreover, the high saturation magnetization, low anisotropy and damping constant of the Py nanodots, together with well-defined magnetic resonance spectra, open a path to their applications in microwave devices based on the fast magnetization dynamics.

## Methods:

**Sample growth.** The sample was fabricated through Hole-mask Colloidal Lithography[24]. It involves three steps: first, a short-range-ordered polystyrene (PS) nanophere array is formed onto a poly methyl methacrylate (PMMA) polymeric layer. Subsequently, titanium is sputtered onto the system before the spheres are peeled-off and the exposed PMMA etched in oxygen plasma, leaving an array of holes as a template for the nanodot growth. Finally, permalloy is sputtered with a DC magnetron sputtering to fill the pores and grow the nanodots, while the PMMA template is removed with acetone, remaining only the nanodots onto the silicon substrate.

**Imaging of skyrmions.** The MFM measurements were performed at ambient conditions using a scanning force microscope from Nanotec Electronica in the amplitude modulation mode and with the phase-locked loop (PLL) enabled to track the resonance frequency. A module for magnetic field application is enabled.[30] Different sensors were used depending on the experiment. Most of them were performed with commercial PPP-MFMR model tips commercialized by Nanosensors[TM].

The non-commercial tips were fabricated following two different procedures. For low moment tips, magnetic cylindrical nanorods were grown onto the apex of AFM tips by FEBID. The 3D magnetic nanowire was fabricated in the commercial Helios Nanolab 600 Dual Beam system equipped with a Schottky field emission gun (S-FEG) electron column and a gas injector system for depositing Fe using $Fe_2(CO)_9$ precursor gas. The nanostructure was grown using an electron-beam voltage of 30 kV, an electron beam current of 43 pA and a chamber growth pressure of $8 \times 10^{-6}$ mbar (base pressure of $1.3 \times 10^{-6}$ mbar), scanning a single point pattern during 75 s with the electron beam. Intermediate moment tips were coated in a home-made built RF sputtering chamber. Co was deposited on tips under $10^{-2}$ mbar argon pressure, with a bias voltage of 295 V and 50 kW of power. The out-of-plane signal of the tips was evaluated with a reference sample. The output, obtained in Hz after enabling the PLL, was correlated to the magnetic force gradient through the following formula:

$$\Delta\omega \approx -\frac{\omega_0}{2k}\frac{dF_{t-s}}{dz}$$

where k is the cantilever force constant and $F_{s-t}$ is the force experienced by the tip as a result of the stray field of the sample.

HRTEM imaging was carried out in an FEI Titan Cube 60-300 system operated at 300 kV and fitted with an S-FEG and a CETCOR aberration corrector for the objective lens from CEOS, obtaining a point resolution below 1 angstrom. It is also equipped with a 2K x 2K Ultrascan CCD camera from Gatan.

**Energy calculations.** To calculate the magnetic energy of the skyrmions in circular soft magnetic dots we consider a dot with the radius $R$ and thickness $L$ and parameterize the unit magnetization vector by the spherical angles, $\mathbf{m} = \mathbf{m}(\Theta, \Phi)$. The angles $\Theta, \Phi$ are functions of the radius vector $\mathbf{r} = (\rho, \varphi, z)$ represented by cylindrical coordinates. The total magnetic energy functional is $E[\mathbf{m}] = \int dV \varepsilon(\mathbf{m})$, with the energy density $\varepsilon(\mathbf{m}) = A(\nabla \mathbf{m})^2 + \varepsilon_m(\mathbf{m})$, where $A$ is the exchange stiffness constant, and $\varepsilon_m$ is the magnetostatic energy.[11] We assume that the skyrmion equilibrium configuration does not depend on the thickness coordinate z and is radially symmetric, i.e., $\Theta = \Theta(\rho)$, $\Phi = \varphi + \varphi_0$ ($\varphi_0 = 0, \pi$ for the Neel skyrmions and $\varphi_0 = \pm\pi/2$ for the Bloch skyrmions). Then, the total magnetic energy as a functional of the skyrmion magnetization is represented by the polar angle $\Theta(\rho)$, $E = E[\Theta(\rho)]$. The energy in units of $\mu_0 M_s^2 V$ ($V = \pi R^2 L$ is the dot volume) is

$$e[\Theta(r)] = \frac{l_{ex}^2}{R^2}\int_0^1 dr\, r\left[(\Theta'_r)^2 + \frac{1}{r^2}\sin^2\Theta\right] + e_m[\Theta(r)], \qquad (1)$$

where $r = \rho/R$, and $l_{ex} = \sqrt{2A/\mu_0 M_s^2}$ is the exchange length.

We define the skyrmion radius $R_s < R$ by the equation, $\Theta(R_s) = \pi/2$, and the reduced radius is $r_s = R_s/R$. The magnetostatic energy density can be written as a functional

$$e_m[\Theta(r)] = \frac{1}{\beta}\int_0^\infty dk\,(1 - \exp(-\beta k))I_z^2(k) + \int_0^\infty dk\, k f(\beta k) I_\rho^2(k), \qquad (2)$$

where $\beta = L/R$ is the dot aspect ratio, $f(x) = 1 - (1 - \exp(-x))/x$, and $I_z(k) = \int_0^1 dr\, r J_0(kr) m_z(r)$, $I_\rho(k) = \int_0^1 dr\, r J_1(kr) m_\rho(r)$, $J_n(x)$ are the Bessel functions of the first kind, $m_z(r) = \cos\Theta(r)$, and $m_\rho(r) = \sin\Theta(r)\cos(\varphi_0)$.

The first term in Eq. (2) accounts for the magnetic energy of the face charges at the dot top/bottom surfaces. The second term corresponds to extra energy related to the volume and side surface magnetic charges. Only the first term in Eq. (2) contributes to the magnetostatic energy of the Bloch-skyrmions, whereas both terms contribute to the energy of the Neel-skyrmions. In this way, the Neel skyrmion magnetostatic energy is always higher than the Bloch skyrmion energy for the same dot magnetic parameters and a finite dot thickness. Introducing a trial function (skyrmion ansatz) $\Theta(\rho)$ in the energy functional (1), one can get the energy of the skyrmion configuration as a function of limited number of the parameters. We use magnetic skyrmion ansatz corresponding to an exact solution of the 2D exchange model by Belavin-Polyakov[41] $\tan\Theta(r)/2 = r/r_s$. This ansatz is a good approximation for soft magnetic dots (no magnetic anisotropy) with relatively wide domain walls. It allows calculating explicitly the skyrmion energy and equilibrium radius $r_s$. The calculated energies of the Neel and Bloch skyrmions are presented in Figure 3b. Both skyrmions are metastable because their energies are higher than the energy of the in-plane single domain state $e[\pi/2] = 0$. The energy of the Neel skyrmions is always higher than energy of the Bloch skyrmions at finite $\beta$ due to the magnetostatic energy of the volume magnetic charges. The calculated Neel skyrmion configurations in circular soft magnetic nanodots are metastable or unstable. The area of metastability of the Neel skyrmion in terms of the dot geometrical parameters is presented in Figure 3a. The Neel skyrmions are the high energy metastable states for the large dot radii $R/l_{ex} > 4 - 6$, and relatively small dot aspect ratio $\beta < 0.8 - 0.9$. The magnetostatic energy increases with $\beta$ increasing, eventually making the Neel skyrmion unstable. The exchange energy increases with the dot radius $R$ decreasing at fixed $\beta$, making the Neel skyrmion unstable at small $R$.

**Micromagnetic simulations.** Simulations were performed with the Object Oriented Micromagnetic Simulation code.[42] The nanoparticles were discretized with a 1x1x1 nm$^3$ mesh and the magnetic parameters used for permalloy were A= 11 pJ/m, $\mu_0 M_S$= 1 T and zero uniaxial magnetic anisotropy. A configuration close to a hedgehog skyrmion was used as initial magnetic configuration.

## Acknowledgement

M.J., E.B. and A.A. acknowledge the support from the Spanish Ministerio de Economia y Competitividad (MINECO) under projects no.MAT2015-73775-JIN and MAT2016-76824-C3-1-R.

M.G-G and A.G-A acknowledge the financial support from the Spanish MINECO project MAT2014-55049-C2-1-R.

J.P.-N., C.M. and J.M.T. acknowledge financial support from the Spanish Ministry of Economy and Competitiveness through Projects MAT2014-51982-C2, MAT2015-69725-REDT and MAT2017-82970-C2, from the COST Project CELINA and from the Aragon Regional Government (Construyendo Europa desde Aragón) through Project E26 with European Social Fund funding. A grant to J.P.-N. was funded by the Ayuda para Contratos Predoctorales para la Formación de Doctores, Convocatoria Res. 05/06/15 (BOE 12/06/15) of the Secretaría de Estado de Investigación, Desarrollo e Innovación in the Subprograma Estatal de Formación of the Spanish Ministry of Economy and Competitiveness with the participation of the European Social Fund.

K.G. acknowledges support by IKERBASQUE (the Basque Science Foundation). The work of K.G. and O.C.-F. was supported by the Spanish Ministry of Economy and Competitiveness under the project FIS2016–78591-C3-3-R.

## Author information

**Affiliations**

*ICMM- CSIC, 28049 Madrid, Spain*
Eider Berganza, Miriam Jaafar, Oksana Chubykalo-Fesenko and Agustina Asenjo

*University of California, Berkeley, California, United States*
Maite Goiriena-Goikoetxea

*University of the Basque Country (UPV/EHU), Spain.*
Maite Goiriena-Goikoetxea, Alfredo García-Arribas, Konstantin Gusliyenko

*Universidad de Zaragoza-CSIC, 50018 Zaragoza, Spain.*
Javier Pablo-Navarro, Cesar Magén, José María de Teresa

*BCMaterials, Parque Tecnológico de Bizkaia, Building 500, Derio, Spain.*
Maite Goiriena-Goikoetxea, Alfredo García-Arribas

*IKERBASQUE, The Basque Foundation for Science, 48013 Bilbao, Spain*
Kontyantyn Gusliyenko

**Contributions**
E.B., M.J. and A.A. conceived the project. M.G-G. and A.G-A. fabricated the samples. E.B. and M.J. performed the MFM measurements. E.B. carried out the micromagnetic simulations with the help of O.C-F.
E.B. and J.P.N. fabricated the MFM tips. J.P.N. performed the TEM measurements. J.M.D.T. and C.M. supervised the growth of the MFM tips by FEBID, the lamellae preparation by FIB-SEM and the TEM experiments. K.G. conducted the analytical study. E.B., M.J. O.C-F. and A.A. wrote the manuscript. All the authors discussed and commented the manuscript.

## Competing financial interests
The authors declare no competing financial interests.

## Corresponding author
Correspondence to Eider Berganza and Miriam Jaafar.

**Figures**

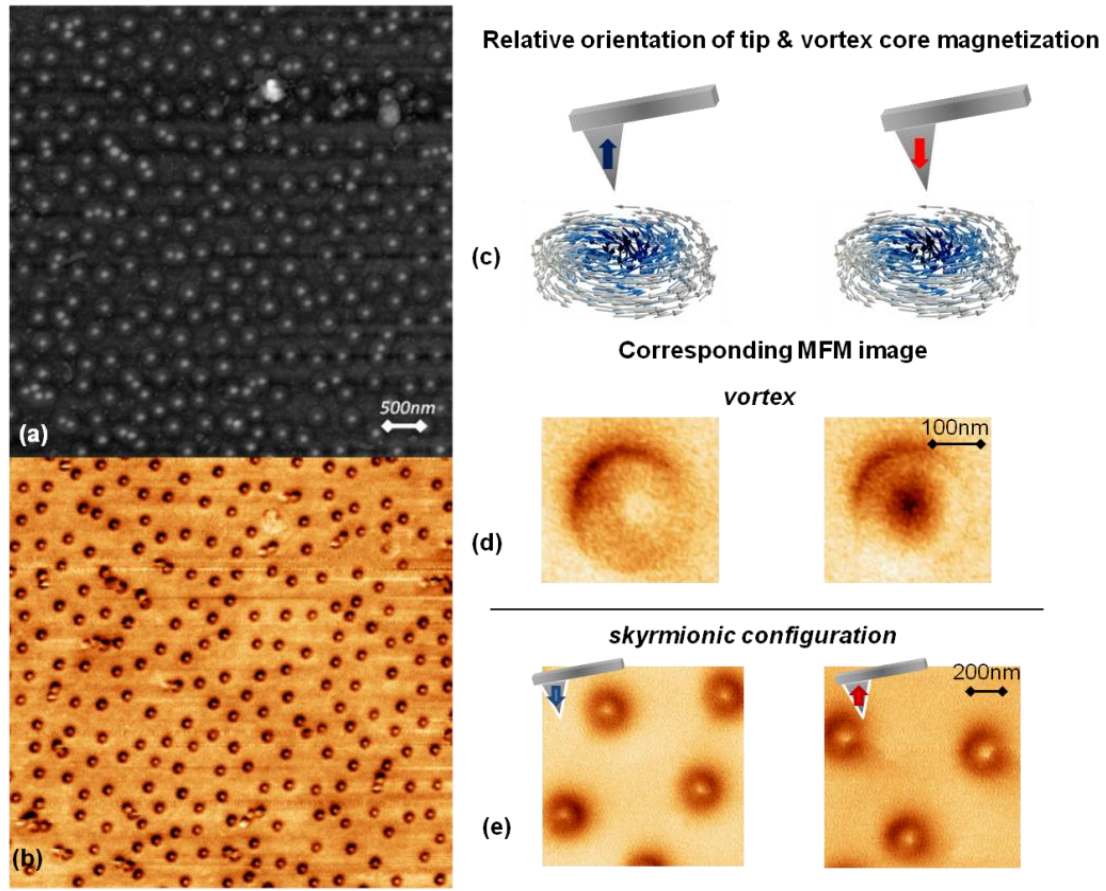

**Figure 1. Dismissing vortex formation.** (a) Topography image of sample A showing the distribution of the permalloy nanoparticles on the silicon substrate. (b) Magnetic Force Microscopy image of permalloy nanodots displaying white cores and dark surrounding areas. (c) Two possible configurations of the MFM tip giving rise to a different MFM contrast for a vortex with the same polarity. (d) MFM images of Py cylindrical nanodiscs with defined vortex configurations (140 nm diameter and 50 nm thickness), where the opposite MFM contrast in the core is only due to an opposite magnetization configuration of the tip. (e) MFM images of sub-100 nm nanodots performed with different tip polarities where the magnetic contrast remains the unchanged.

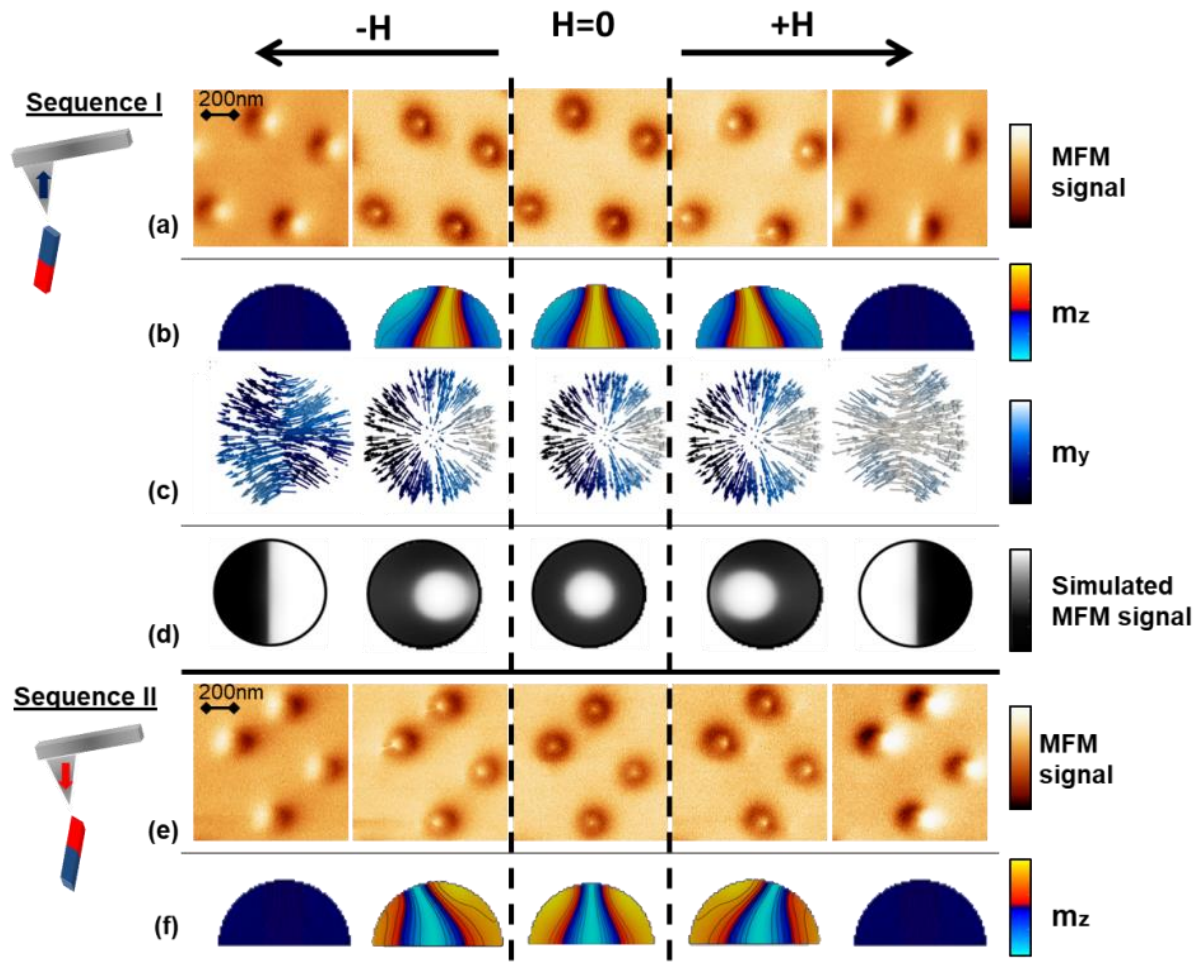

**Figure 2. Skyrmion evolution under in-plane magnetic field.** (a) and (e) show schematically the defined state of the MFM tip polarity prior to sequence I and II, respectively. MFM images in (a) and (e) show the evolution of the core of the skyrmion moving antiparallel (parallel) to the applied in-plane magnetic field. ±20mT are applied to displace the core from the centre and ±55mT to saturate the nanodots. Simulated hedgehog skyrmions of positive and negative core polarities (b) & (f) reproduce the displacement of the skyrmion core in sequence I and II. In (c) the transverse configuration of the hemisphere base displays a radial arrangement of the in-plane magnetization in hedgehog skyrmion of positive polarity. The simulated MFM images of sequence I are presented in (d).

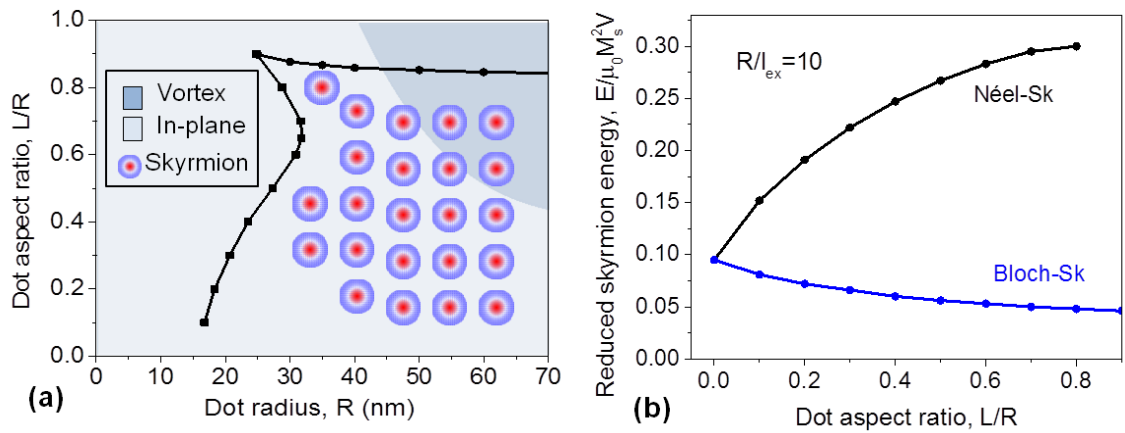

**Figure 3. Phase diagram.** (a) Phase diagram indicating the region of the dot sizes where the hedgehog skyrmion exists in permalloy nanodots. The black dots and lines correspond to the boundaries for the existence of a metastable skyrmion configuration according to the analytical calculations. The light blue and the grey regions denote the ground states corresponding to in-plane single domain and vortex configurations, respectively. The red-blue dots mark the points where (meta) stable skyrmion configurations were found by micromagnetic simulations in hemispherical dots. (b) Energies of Néel and Bloch (vortex) skyrmions as a function of the dot aspect ratio as calculated analytically. $l_{ex}$ is the material exchange length.

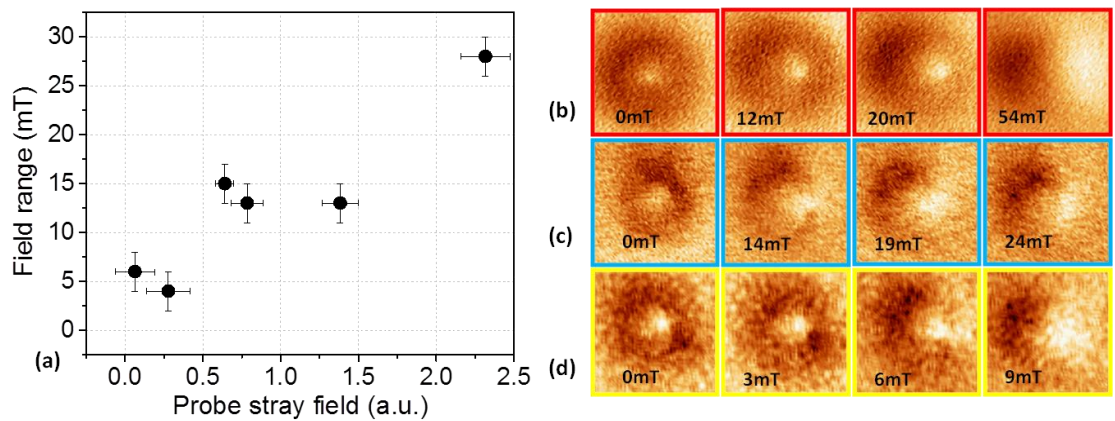

**Figure 4. Determination of skyrmion stability as a function of the magnetic force gradient.** (a) Correlation between OOP magnetic stray field and stability of the skyrmions. The black dots represent the experimental data obtained from the MFM based measurements described in Supplementary Information IV. Field sequences performed with different tips show decreasing saturating fields from (b)-(d). Image size 250x250 nm.